\documentstyle[prl,epsf,epsfig,aps]{revtex}

\newcommand{\myskip}[1]{}

\newcommand{\BEQ}{\begin{eqnarray}}
\newcommand{\EEQ}{\end{eqnarray}}
\newcommand{\BEA}{\begin{eqnarray}}
\newcommand{\EEA}{\end{eqnarray}}

\title{Round Table Discussion at the Workshop ``New Directions in Modern Cosmology''}

\author{Theo M. Nieuwenhuizen$^{1,2)}$,  Peter D. Keefe$^{3)}$ and Vaclav \v{S}pi\v{c}ka$^{4)}$}
\address{
$^{1}$ Center for Cosmology and Particle Physics, New York University,  4 Washington Place, New York, NY 10003, USA \\
$^{2}$On leave of: Institute for Theoretical Physics, Science Park 904, P.O. Box 94485, 1090 GL  Amsterdam, The Netherlands\\
$^3$University of Detroit Mercy, 24405 Gratiot Avenue, Eastpointe, MI 48021,USA\\
$^4$ Institute of Physics, Academy of Sciences of the Czech Republic, Na Slovance 2, 182 21 Praha 8, Czech Republic}

\begin{document}
\maketitle

\begin{abstract}
The workshop ``New directions in modern cosmology'', organized by Theo Nieuwenhuizen, Rudy Schild, Francesco Sylos Labini and 
Ruth Durrer,  was held from September 27 until October 1, 2010,  in the Lorentz Center in Leiden, the Netherlands. 
A transcript of the final round table discussion,  chaired by  Theo Nieuwenhuizen and Rudy Schild,
is presented. The subjects are: 0) spread in data; 1) back reaction; 2) $N$-body simulations;  3) neutrinos as the dark matter; 
4) gravitational hydrodynamics, 5) missing baryons and lensing in an inhomogeneous universe,  and 6) final points. 
\end{abstract}

\section*{Introduction}

{\it Theo Nieuwenhuizen:}
		There are a couple of subjects that I would like to bring in for this discussion. Point 0 is the spread in the data themselves. Point 1 will be about dark energy or back reaction or inhomogeneity, we have heard 3 talks about that.  Point 2 will be about $N$-body simulations and dwarf galaxies and, as it was announced yesterday, no $\Lambda$CDM.  In point 3, I will break in for 2 or 3 minutes saying something about neutrinos as dark matter candidate, and then discuss that scenario.  And for point 4, we go back to the first talk of this morning, gravitational hydrodynamics or hydro gravitational dynamics or baryonic dark matter, whatever you want to call it.  If we have time left over, we will get to point 5 and 6.

\section*{SUBJECT 0: SPREAD IN THE DATA}
{\it Theo Nieuwenhuizen:}
Point 0 is of course obvious.  We have the data and that is what we should describe.  In particular, there is noise in data.  Of course some part is statistical.  You don't want to care about that.  But let us keep in mind that some part can be non-statistical and can be part of our physics.  We have seen this in the CMB plot this morning where you had enormous spread of individual data points of the temperature fluctuations and it can be well that there's a lot of physics.  That you do not only describe the average which is one property, but there's also a variance and higher moments and so on. 

	Let us keep that in mind that there can be important physics both in this CMB spread of individual data points and what we saw later for the supernovas.  There was also an enormous spread.  Of course, the more data points we have, the better average we get.  We should describe  not only the average.  In the end, we should describe the whole data set and it could be more physics than we are now aware of.  

{\it Francesco Sylos Labini:}
	 One of the main things I have learned here is that we have to take care of systematics much more.  Knowing the error bars that we see in this plot, may be completely meaningless if the systematics have not been taken into account. So to do that is a hard job. If there is somebody that is willing to invest some time in understanding the data then that is something very useful.Ó 

{\it Theo Nieuwenhuizen: }
	ÒThank you, Francesco. Yes we have seen one person, Richard Lieu, who had spent a lot of time on WMAP data. Thank you for your point.Ó

{\it Alexander Unzicker: }
	ÒI am advocating that the source code of data processing should be as public as possible. That everybody should try, at least try to be as reproductive as possible.  I think weÕre doing this in the wrong way because everybody works in these codes and it's not really transparent.  This is a big problem and everybody should think about it.Ó
 
{\it Theo Nieuwenhuizen:} 
Indeed, it may save us from being on the wrong track for decades.

\section*{SUBJECT 1: BACKREACTION}

{\it David Wiltshire: (short intro)}
   	ÒThere are various approaches to dealing with models of the universe. Traditionally we use exact solutions and CM homogeneity and get the Friedman model and the $\Lambda$CDM model which is really a Friedman model plus tiny perturbations, so this is the standard cosmology.  We've also heard about you can give up the assumption of homogeneity but have isotropy and you're led to these large void models so, I'm not sure whether one of the questions is how it needs to be treated.  Chris Clarkson said they're crazy even according to many people who use them, so what do they tell us.  Independently of that, there is a test of the Friedman equation which is of course put forward, so this homogeneity assumption is actually testable but to actually do it is going to require measurements of $H(z)$  and that might require something like the Sandage load test of the red shift drift which we maybe deal with the next generation of extremely large telescopes.Ó   

	ÒSo, we can certainly test that standard paradigm with enough technology.  So we go to inhomogeneous, things which are not exact, but of course first we have a Friedman model with Newtonian perturbation, this is a standard cosmology so one question is how much does this agree or disagree with observations.  That was the question that Ruth posed in the first round table discussion of this workshop.Ó

{\it Ruth Durrer: }
	ÒI have a question: what happens on small scales ?Ó

{\it David Wiltshire:}
	ÒCorrect.  So, these perturbations ... what I was thinking here mainly was in terms of the $N$-body simulations.  $N$-body simulations pose results.  Certainly, general perturbations you can treat by many different means at the surface of last scattering; here I'm particularly thinking of structure, large scale structure and what's currently used. In this sort of framework there are discussions about back reactions in the perturbative regime.  This is the arguments that went on a few years ago between Robbie Kolb and others.  We haven't had a lot of discussion about that at this meeting.  So that's one route you can take.  Here, what Alexander Wiegand and I both decided to look at was under-average evolution and e.g. the Buchert  approach and then you've got a question: is it just the average or is the fitting problem also important and this is where I diverge from Thomas Buchert in that I think the fitting problem is also important in trying to address that and make a testable and falsifiable model.  I don't know what people see as the biggest questions here ... I wasn't asked to pose a question, but this is the scheme of things we've been discussing so ... Ó 

{\it Theo Nieuwenhuizen: }
	ÒSo let's take questions.Ó 

{\it Jan Smit:}
	ÒIn the Buchert scheme you have a hyper surface chosen only one time for the hyper surface.  Now, you're saying somehow that's wrong.  Why is it not just a gauge choice?Ó 

{\it David Wiltshire:}
	ÒWell, what I'm saying in my approach is that indeed it is a gauge choice and you shouldn't restrict yourself to having a single gauge choice.  I'm looking at different gauges and trying to mesh them together somehow.   So I'm saying that even though in the Buchert schemes, things are expanding at different rates.  That there is a choice of slicing in which the expansion is locally uniform and itÕs actually general relativity anything which is related to first derivatives of the metric ... that's the connection.  Connections are, in a sense, a gauge choice and the Hubble parameter is, in a sense, a gauge choice.  We should be able to do our averaging and think about these problems in ways in which we understand when we make a gauge choice and what things should be independent of gauge choices.  The way that I'm looking at it is that those gauge choices are important in the fitting problem.Ó

{\it Jan Smit:}
	ÒBut then the connection to how you want to use it to connect these observations comes into play.  Isn't that the essential issue?Ó

{\it David Wiltshire:}
	ÒYes.  So, my approach to this essential issue, because I do it by Ansatz,  which I think is well motivated, so one would say that the difference between any other homogeneous model of this sort and the Friedman model is that in the Friedman model we have a uniform expansion then everything has a velocity perturbation on top of it, whereas here, what you think of as a peculiar velocity is in some sense actually most of it is due to variations in expansion because of regions of different density.  So, the velocity perturbations around any scale are going to be small.  Then there's a gauge choice about how you relate these different things.  I know the weak point of my approach, as far as mathematical relativist are concerned, is that I haven't solved that problem to the level of satisfaction that people would like.Ó

{\it Francesco Sylos Labini:}
	ÒMy question is not a question, just a comment in general.  In this approach when I see the papers ...   ...  people trying to include different matter distribution then just the uniform density field.  But then, for example, especially from Thomas Buchert models, I can see an effort to go into the data and see whether the actual distribution of galaxies fit with the theoretical prediction of your models.  In your case, basically, what you are interested in is the size of the voids.  How much matter, how big are the voids.  The parameter you want to get from the observations and use for the theoretical predictions.  While in the  Lemaitre Bondi model, I don't see something equivalent to that.  For example, I was trying to develop a test to see the effect of the void but on scales which may not be relevant for what you do.  If we are in the center of the void, then we look far away, we have to see a change in the density according to the density outside the void itself.Ó

{\it Chris Clarkson:}
	ÒIt's just a function of scale.  ItÕs going toward 100 mega parsecs.  The density profile in the very large void models just doesn't change much.Ó 

{\it Francesco Sylos Labini:}
	ÒWhat about the fluctuations.  Which level of fluctuations do you expect or can be compatible on some scales?Ó

{\it  Chris Clarkson: }
	ÒSo it's an open question where perturbations are involved, but along the central world line one can assume that they evolve as in a curved Friedman model.  The trouble is, you don't have an amplitude to compare it to with your primordial power spectrum from the CMB because you don't know the initial amplitude.  You don't have another observation for that.  So your $\sigma_8$ is just a measurement of your primordial density power along the central world line.  So a priority it's difficult to know what that  could tell you as a constraint on the models rather than just an observation of a particular part.Ó 

{\it  Theo Nieuwenhuizen: }
  	ÒDo I collect more questions?  Yes.Ó 

{\it  Giovanni Marozzi: }
	ÒWhen you say a gauge choice, do you mean the dependence on the observer?

{\it David Wiltshire:}
	ÒIf spatial curvature isn't the same everywhere, (in the Friedman model we had that thin ... ) The question is: If I've got large variations in spatial curvature, how do I calibrate rods and clocks of some average statistical quantity with those of any particular observer.  So when you have genuine inhomogeneity, there is variance in the geometry.  So, not every observer is the same observer.  That's what I think is the problem with this approach in that we just forget the fitting problem.  Here we just pretend the universe is a smooth fluid which is homogeneous isotropic.  When we pretend it's a smooth fluid that has got spherical symmetry, but it's still a smooth fluid.  We don't ask: what are the fluctuations and what is the result of having big fluctuations on a particular scale.  For me, the fitting problem is very important and the universe gives a natural observer selection effect.

{\it Giovanni Marozzi:}
	ÒDid that correspond in a formal way with the observer.  Did it not connect different observers like with a mathematical function.  Cannot you define away with the observer?Ó

{\it David Wiltshire:}
	ÒIf you've got an exact solution or, even a ÒSwiss cheeseÓ model (in a Swiss cheese model you punch holes in the cheese; the  evolution so, in a sense, it's still Friedman at some level) then everything is well defined because you just write down your matrix and you've got a single matrix and everythingÕs Òhunky-doryÓ.  The problem comes when we're observers and super nova and distant galaxies are effectively things that are being observed.  So, in between, there is nothing in the void.  So, we may have a messed biased view of the universe.  So, the question is:  If I've got some average quantity ... if I'm doing something like a spatial average on some big cube, we can always define statistical things.  How do those statistical things relate to the observations of the particular observers in the galaxy?  So, what I'm suggesting with the fitting problem is that the way those statistical quantities relate to the actual observations of observers in galaxies, don't have to be the same as the way that they might relate to some hypothetical observer who probably doesn't exist in actual reality and who would be not bound to any structure but in a void.Ó 

\subsection*{SUBJECT 2: $N$-BODY SIMULATIONS}

{\it Theo Nieuwenhuizen: }
ÒSo, it will be about the exciting talk that we heard yesterday.  I could hear and see a big silence in the audience yesterday afternoon which, of course, if you give a talk and there's a big silence, you always have to watch the audience to see whatÕs going on.  We had such an occasion here yesterday, not due to the boring talk, flabbergasting was what was going on and let us recall shortly what was the issue and then discuss it.Ó

{\it Pavel Kroupa:}
	ÒBasically, I was asked to put up 2 slides maximum.  I have put 6 together into 1 slide and just yesterday, I tried the argument with the conclusion essentially and that is that, if this is true in whatever framework, even if you believe in the planets or if these are tidal dwarfs in an expanding tidal tail, then one is forced to essentially conclude that we are in this sort of situation the real universe where purely baryonic galaxies merge and produce dwarf galaxies surrounding the larger galaxies and given that then, this is the statement which is correct.Ó

{\it  Francesco Sylos Labini:}
	ÒThere is a question that I wanted to ask you yesterday.  Basically, your conclusion is that the simulations don't fit the observations.  You can say that $\Lambda$CDM doesn't work because you are comparing the result of the simulation with the result of the observations.  This is again a question of systematic effects.  The problem is up to which scale that was the problem addressed yesterday in the first lecture, in the talk before yours, Michael, so up to which extent you can think that the simulations are really reproducing the models because, in the simulations there are a lot of parameters which are not in the model.  How much they influence the numerical evolution of the particles is not really clear and it's an open question because, again, what is missing is a systematic studies of effect of these  discretizations of dynamical  evolution of  the real system.  I mean, it is not like their predictions of the model  ...  but mixed with a grain of salt.Ó

{\it  Alexander Unziger:}
	ÒIf you do that then you are left without the testable theory and then it's metaphysics. For me science is, you have a theoretical framework, you make predictions, then work all that out.  This is what this is. I'm from the numerical sector myself so, I can also estimate whether the predictions are reliable following the advance in code technology, as well as the computer architecture and just generally the advancing in computer capabilities over the last 15 years.  The situation is very robust.  It's not so that you can change some parameters in the $\Lambda$CDM framework and get completely different results concerning the dark matter substructure this is completely ruled out but if you claim that is the case, then one would have show it in a refereed publication and, until that is the case, I think one has to simply look at the fact that this is the situation.Ó

{\it  Francesco Sylos Labini:}
	ÒI don't object that this is the situation, the problem is more conceptual.  The problem is to understand in which limit the simulation of particles is approaching the theoretical model.

{\it Ruth Durrer:}
	ÒI fully agreed with your situation of the observations but, IÕm not sure that this is really ruling out $\Lambda$CDM for the following reason:  Clearly the $N$-body simulations of the $\Lambda$CDM show much too much small scale structure.  On the other hand, we know that on small scales (small, let's say out to roughly one mega parsec maybe even more but roughly the size of a cluster) I would not be surprised if other pieces of physics, which are not in $N$-body simulations, become relevant.  Be that hydrodynamics, turbulence, magnetic fields and all these things are not in the $N$-body simulations, and if they become relevant, then on these small scales the $N$-body simulations are not the relevant thing.  That's why I think it is very difficult to rule out $\Lambda$CDM by looking at small scales.Ó

{\it  Theo Nieuwenhuizen:}
	ÒI think one can never really ever rule anything out, but one can just bring more arguments in favor or against.

{\it Pavel Kroupa:}
	ÒWell, personally I think one can rule out.  I think that stage is already achieved.Ó 

{\it Alexander Unzicker:}
	ÒI would agree with Francesco but, this is not a counter argument to Pavel Kroupa, I think it's the business of the people who do the dark matter simulations to comment that right.  I mean, the fear could be that extra simulations with extra parameters, one day could fit even the anomalies you are pointing out.  But, I would be careful to encourage you people to introduce more parameters to just  fit. Even if it's called a physical process like magnetic fields because at the very end we know very little and the more parameters you put in, the easier it would fit.  By the way, I think he showed, in a very general way that the very basic concept of halos is in contradiction.  That's, I think, quite model independent of how you do the simulations.  He showed that the very idea of a dark halo is in contradiction with many observations and that will be hard to fix. 

{\it Carl Gibson: }
	ÒI think you really did rule out halos. I agreed with your conclusion.Ó 

{\it Luciano Pietronero:}
	ÒI'm a little embarrassed to hear that this $\Lambda$CDM cannot be ruled out or can be ruled out.  It looks to me that we are not in presence of a theory.  We are in presence of a filthy scheme in which we have a large number of parameters and a large number of assumptions and at any moment we are allowed to pick new effects,  scale corrections bias and so on.  And out of this incredible soup, then we say nothing is excluded but, if this is so, Ruth, this means nothing.  This contains zero information so the information exists only if you get into something very clearly.  Otherwise, if it doesn't fit, you  add bias.  You add this, you add that and you define bias as the difference between experiment and theory.  As dark matter is also the difference between experiment and theory.  So, I would like to push the discussion into something which is absolutely outside the prediction of whatever this blob is.  Otherwise, it's useless. 

{\it Ruth Durrer:}
	ÒJust a quick answer to why I think on small scales, the physics is too complicated.  There are too many effects that you cannot really use to rule out $\Lambda$CDM.  So go to someplace where the physics is more simple.Ó

{\it Luciano Pietronero:}
	ÒYes, but you do not answer my question.  What do you exclude, Ruth?  Ó

{\it Ruth Durrer: }
	ÒFor example ... If you would measure the deflection of light around the sun, would be twice the solar potential.  You would have 
	ruled out general relativity.  This measurement has been done.  It's a simple thing.  It could be calculated. One knew what one has
	 to get ... etc.  The $\Lambda$CDM, I think we also have some pieces of simple physics, and the simplest, I think, is the CMB.  And if we could rule it out there then that would be clear cut.  But on small scales, it's difficult.  Maybe with large scale structures we can also do it with the 100 mega parsec, 200 mega parsec.Ó 

{\it Rudy Schild: }
	ÒI think that astronomers like to take refuge in the CMB which we can analyze and reanalyze forever.  I think that Dr. Kroupa's point, that we do not see these objects which are a fundamental, is a profound prediction, yes, prediction, now of the cold dark matter theory. It is the death now of cold dark matter theory.  I am extremely impatient of the idea of adding an epicycle to that theory.  That was discarded hundreds of years ago.  In other words, going back now and saying we add a new parameter to our theory to have these sub-halos disappear in the local universe is adding an epicycle.  That's a posteriori calculation.  I think that science should get away from that. 

{\it Richard Lieu:}
	ÒI should add that Simon White did email me two or three years ago, pointing out that this is in his opinion the single 
	 biggest blow to $\Lambda$CDM.Ó

{\it Pavel Kroupa:}
	ÒConsidering small scale, I hear this a lot.  It's always said whenever all of this is brought up it is always said Òyes, we don't understand the small scale physicsÓ.  Now, maybe the reason why I'm so completely convinced of the essential of this statement is because I come from the small scale physics.  I started my research career by looking at the nearest star, Proxima Centauri, and then advanced to understand star clusters and stellar dynamics and how stars form out of gas.  I'm very well acquainted with that and I know what that are and I know what dynamics is about concerning galaxies.  My conclusion is that the uncertainties ... there are uncertainties but they're by far not of that magnitude to impact on that argument.  And that's why I'm making the argument because I've convinced myself that that is the situation otherwise I would not be standing here making this statement which goes against 99.999\% of the astronomical community.Ó

{\it Theo Nieuwenhuizen:}
ÒNot that much of th\'is communityÓ.

{\it David Wiltshire:}
	ÒSo, strong and weak gravitational lensing is a very powerful argument for the standard picture.  Quite apart from the cosmic micro background radiation, I think that gravitational lensing is something that is very hard to get right in some other model.  Something with epicycles like TeVeS if you want to do it and in model scenario for example.Ó

{\it Pavel Kroupa:}
	ÒWhy would that be an epicycle?Ó

{\it  David Wiltshire:}
	ÒWell, you're trying to get a whole theory of gravitation.Ó

{\it  Pavel Kroupa:}
	ÒThat's not an epicycle.Ó

{\it David Wiltshire:}
	ÒSpeaking as someone who comes from a background of theory. I did continents and many other things and I'm a theorist who's used to adding what I think are theoretical epicycles which are not particularly well motivated.  It's very easy to add a field that you've never observed just for the purpose of fitting rotation curves.  I think that if you're doing that with TeVeS then effectively your adding epicycles in a way that you add epicycles in other places. The MOND  phenomenology is really interesting but it's saying something which is much deeper I think than simply adding a field to the action. 

{\it Pavel Kroupa:}
	ÒIf I might just add, consistency of the theory ... how much you love the theory ... as much as you might love the theory, is not a proof of the theory.  That's how we exclude, actually,  models,  by looking at the disagreements and that what has been done here so you might find agreement in certain range of parameter space from large substructure but it doesn't prove the theories.Ó 

{\it David Wiltshire:}
	ÒCertainly but, there are so many theories because itÕs very easy to cook up certain theories and the theory has to be falsifiable.  But amongst all the theories that you can choose to examine because, there is an infinite number of them, you do so on esthetic grounds.  Inflation, there are a few hundred models of inflation so, I think inflation is something more like phenomological thing that's easy to get though we don't really understand it at a fundamental level.  So, I agree with you that you can have parameters and it should be falsifiable.  Many of these modified gravity theories have problems at the solar system level.  So, the first thing is, can you rule it out for other reasons.  So, it's a judgment in saying that it's epicycles.  I think the phenomenology is interesting, in saying something that we probably missed some deep point.

{\it Ruth Durrer: }
	ÒI would like to add something about  extra's  which you need to make flat rotation curves.  If you want to compute the CMB with  something extra, you have to have a realistic version of it and there's not a realistic version of MOND, the only one I know is  TeVeS and if you want to get the CMB correct with  TeVeS you have to introduce massive neutrinos, which is ok, it's not so bad, right.  So you have dark matter but you also need $\Lambda$.  You cannot get the CMB correct without $\Lambda$. You introduce a cosmology,  you fix your parameter which you have to get the solar system correct to get the flat rotation curves and to get the MOND like dynamics on galactic scales and I think you need ... you also must make careful not to get the problems which you can't do with usual MOND. And then, if you want to get the CMB also right, you have to introduce $\Lambda$.

{\it Pavel Kroupa:}
	ÒI'm not arguing for MOND.  I'm saying $\Lambda$CDM doesn't work. I mean what are the physicists and now, look at your research budgets.  99\% of it goes into $\Lambda$CDM research.  There are maybe 3 people in the looking at MOND in a cosmological context.  So, you can't say MOND fails. You don't have the large quantity of theoretical as well as observational results to do testing.  You can't do it. 

{\it Ruth Durrer: }
	ÒIf you had looked at the TeVeS model, maybe you would say, ok, that's epicycles.Ó

{\it  Pavel Kroupa:} 
	ÒIÕm not arguing for TeVeS, I'm saying $\Lambda$CDM doesn't work.  Now the conclusion is, we have to do a lot of work to find an alternative.  Is general relativity wrong or right, I do not know.  It might be right; It might be wrong.  Maybe there are additional effects in the field equations, which we haven't dealt with. We discovered a new effect which was unknown 5 years ago.  Possibly the back reaction  ...   Write your first paper, but that's a very new effect.Ó 

{\it Pierre Astier:} 
	ÒI just wanted to say that the points of failure  are linked with bound states of a gravitationally bound  system,  which is something complicated.
	 We have another example which is QCD.  We never threw away QCD because we not able to compute bound states.  And, that is you can do a theoretical QCD, something which is calculable can be checked.  And it works.  Bound states are extremely difficult and I'm not surprised that in a theory like $\Lambda$CDM we had a hard time in computing bound states.

{\it Pavel Kroupa:}
	ÒYou lost me there.  That's Newtonian dynamics what you're talking about.  I thought that that is understood pretty well.Ó

{\it Pierre Astier:} 
	Ò In Newtonian dynamics, no one is able to solve anything rigorously. It is an $N$-body problem. I'm not surprised that we don't get the right solution. 

{\it Theo Nieuwenhuizen:} 
	ÒWell this is a very hopeful position.  Ahhh ... Francesco, you go first.Ó 

{\it Francesco Sylos Labini:}
	ÒI think the point is that we don't know, nobody knows, is how to have some analytical solutions in the non-linear regime in gravity.  So this is a problem which goes beyond $\Lambda$CDM.  And if you take any initial condition and any cosmology (which is funny because cosmology doesn't change very much in these simulations) you'll always form the same kind of new linear structures which are ... these structures which are called halos. Now the problem is I would say given that what you observed is different from these halos.  Or there are other ingredients, as Ruth was saying, that are missing in the gravitational simulation or maybe there is a problem with the standard way in which gravitational simulations have been done.  And this is something that we are studying.  We think that probably there can be a problem in the numerical simulations.Ó 

{\it Pavel Kroupa: }
	ÒSure,  I would fully accept it if what you are saying showed that there is an effect which one has to fully understand in the nonlinear regime at which effects what galaxies look like in a very, very significant manner and of course, that would be a tremendous break though, of course.Ó 

{\it Francesco Sylos Labini:}
	ÒOn the other hand, the predictions of  $\Lambda$CDM in the linear achievement (so at 100 mega parsecs) then there are no problems in linear regime, there we understand a lot of things, more or less everything.  So, for example:  If the baryon peak is not there, then it's a big problem for $\Lambda$CDM.Ó 

{\it Ruth Durrer: }
	ÒI think this is an important point which now Francesco has brought. Probably this lots of substructures which is in the numerical simulation has nothing to do with $\Lambda$ but more with CDM of course, itÕs just Newtonian clustering of particles with all others.Ó

{\it Pavel Kroupa:}
	ÒSo will we be challenging the cold dark matter aspect of the theory or the $\Lambda$ aspect of the theory?  I mean if you are challenging ... So which of those aspects would you be questioning?Ó

{\it Tom Shanks:}
	ÒIt seems to me that I don't mind the argument that if you push away from CDM you have to want the flat rotation curves.  I just don't think that applies.  If you have a surface density that falls off as $1/r$ in a spiral galaxy that gives you a flat rotation curve and H$_1$ neutral-hydrogen falls off as $1/r$ exactly where you want, by a factor 7 too low,  then if you fill up with molecular hydrogen (which is very difficult to observe) you have to hide it from UV surveys; it has to be in a special fractal. Francoise Combes showed us this a few years ago ... then you use the CO when you see orbservational curves of the galaxy.  That is an H$_2$ tracer and it did seem to be in that fractal distribution that she wanted. So, I don't think you can jump from abandoning CDM to having to have MOND just because of flat rotation curves.Ó

{\it  Ruth Durrer: }
	ÒDo you think one could do it without baryons?Ó

{\it   Tom Shanks:}
	ÒYes, yes because the argument also used to be that you were prone to baryon stabilities and suppose you didn't have anything like that standing spherical halo.  But these days the bars of observation are smaller anyway, and that problem, to some extent, has gone away.  So, flat rotation curves ... I mean itÕs my opinion    ...  can be extremely boring and say there are errors in the observations or things we know anyway, but in this case, I think itÕs true.  Flat rotation curves I don't think equate to satisfaction.  We could also talk.  It's always good to stir up the conference with the bullet cluster, which nobody has mentioned; I'm amazed at this point.  That's a very interesting point too.  Fight usually break out when the bullet cluster is mentioned.  I think that's sort of a swindle of an argument but we could get into that as well.Ó

{\it Carl Gibson:}
	ÒÒI know you can rule things out observationally but, can't you rule out some ideas theoretically. To me from a fluid mechanical point of view, the concept of cold dark matter is absurd.  It's completely inconsistent with the concept of fluid mechanics.  That something with a very low collision cross section could condense.  Alright ...  if you make it absolutely cold dark matter, then I will agree.  If it's absolute zero, you start off with an initial chunk of cold dark matter somehow given to you initially, then it will condense.  But then it will defuse away and the idea that you could have a permanent halo of it is absurd.Ó 

{\it Jan Smit: }
	ÒThe interactions are just very weak.Ó

{\it  Pavel Kroupa: }
	ÒLet me just say something.  So, if you say it's natural, then how does that now agree with that.  With what you've been saying.  Because you're saying that maybe the substructures are not there.  But you just said it's natural, so ... Ó

{\it Theo Nieuwenhuizen: }
	ÒWell, what Carl is pointing is the fluid mechanics of the cold dark matter itself which I have never seen any discussion of it.  We as humans ...    ... Carl is saying, if you look at the fluidity of this stuff, you must convince yourself that it will cluster and stay clustered.Ó 

{\it Carl Gibson:}
	ÒSuppose you had some of that.  What would you put it in? It would diffuse away.  It's like having a bottle of neutrinos.  What would happen, it would diffuse through the walls.Ó 

{\it  Francesco Sylos Labini:}
	ÒIt depends on gravitation. If you have gravitation of  a collisionless  system, that way it works perfectly it won't diffuse away.Ó  If you do the simulations ... 

{\it Carl Gibson:}
	ÒIf these are point particles with a very low cross section for collision, they will diffuse away.Ó

{\it Ruth Durrer: }
	ÒWithout gravity, yes. Ó

{\it Carl Gibson:}
	ÒWith gravity, they will diffuse to the center ... 

{\it Ruth Durrer: }
	ÒNo, not necessarily ...   ... some of them go out into infinity and some of them go into the center.Ó 

{\it Theo Nieuwenhuizen: }
	ÒLet us agree that there is a point here.  And then that's the matter and also ... yesÓ 

	{\it Francesco Sylos Labini:}
	ÒThe problem is that, how certain particles relax to a kind of stationary state.  OK, that's the question.  So, what's the physical mechanism?  In certain systems there are mechanisms which are not based on particle collision.  They have a coherent  effect on those systems what is called violent relaxation this is the way in which ... otherwise we would not be here without violent relaxation.Ó 

{\it Theo Nieuwenhuizen: }
	ÒSo we must check for violent relaxation of the cold dark matter.Ó

{\it Pavel Kroupa:}
	ÒNo, because haven't you just sort of argued against yourself?  Because if you said there's violent relaxation, which I understand pretty well I think, (that's why this is here) then you form these substructures.  I mean, there's no way out of that.  So either you say the simulations are not convincing but then you can't say violent relaxation forms the halos or you say that the simulations are convincing and then you can say that you've got violent relaxation and it formed the halos.  So what is the status?Ó

{\it Francesco Sylos Labini:}
	ÒWhat happens in the simulations is something very difficult to understand.  So, it's not violent relaxation, which is something that has been understood in an isolated system.  The relaxation occurs in an infinite system in which you have contribution to the force from many different scales.  So, it's a process which in unknown actually.  It's something like violent relaxation but it's not the violent relaxation we know for an isolated system because it's not isolated.

{\it Theo Nieuwenhuizen: }
	ÒRudy ... you have a remark?Ó

{\it Rudy Schild: }
	ÒThis really was a question.  Ruth brought up MOND and I think some of you may know some things about MOND and frankly, it was my understanding that when you make the MOND modification which is then the relativistic version which then is applicable and tuned to the WMAP to the SZ effect of weak lensing that then you don't get the correct calculation of strong lensing of quasars.  In other words, the double imaging of a distant quasar or by the foreground galaxy.  Does anybody know the status of that now?  Maybe Ruth?  Maybe another of you?Ó  Thank you, that was just a question.Ó 
	
	\section*{SUBJECT 3: NEUTRINOS AS THE DARK MATTER}
	
{\it Theo Nieuwenhuizen: (short intro)}
	ÒOK, given the time schedule of the discussion, we are doing fairly well.  We are slightly ahead and slightly halfway through the series of topics.  I now give Rudy the chairmanship and I will now become a member of the discussion team.  A short, short story about neutrino dark matter because I have brought it up a few times in this conference. I look at a galaxy cluster, Abel 1689 is well known, and the idea is just to solve the Poisson equations so there's Newtonian physics and the mass density is of course dark matter next to  Hydrogen gas and galaxies.  So, there are 3 ingredients to the matter.  I take an isothermal model, we know this very well, but I  apply it for isothermal fermions.  So they give a Fermi Dirac distribution of free particles in the gravitational well of all this stuff together.  Then there is   X-ray gas and this I take from observations.  Then I get the electron density and proton density and temperature, and so on.  And galaxies I simply model as isotherma;l this is not completely correct but, that's what I did and there was some data. These are lensing data.  This is weak and strong lensing and some data are based on a mixture of weak and strong lensing.  From the figure I show, you see that the fit works.  What comes out is a mass of the putative dark matter particle and an 1.5 electron volt mass would fit the best to neutrinos.  The active neutrinos, the normal ones, would yield 10\% dark matter in the cosmic budget.  So your active ones would be 10\% and the right-handed ones or sterile  ones could also have 1.5 electron volt and in principal they can also be 10\% or maybe even more, but probably depending on the Majorana coupling in the mass matrix.  This would then go together with baryonic dark matter.  So, there would be a 4.5\% baryonic dark matter as we understand it today and there could be 20\% from these neutrinos and at least what this shows is that the cosmology that you get from this is not too bad.  But we could not go to the CMB predictions because it was just started and it would be  an important task to do still and also difficult.  That's the situation for today.  Rudy, I see a question. 

{\it  Rudy Schild: }
	ÒYes, the question in back please and your name?Ó

{\it Pierre Astier: }
	ÒWhat is the temperature of the neutrinos?Ó

{\it Theo Nieuwenhuizen: }
	Ò0.04  Kelvin. It is below the 1.9 K because the 1.9 is the momentum temperature $T_p=\langle| p|\rangle c$, but, what you have to think of here is the kinetic temperature and that's $T_K=\langle p^2\rangle/2m$  and this is well below this 1.95  K.  And the value I find is above this kinetic temperature.  If they would not be clustered on the galaxy cluster which would still be homogenous they would have a few times $10^{ -4}$  K and this 0.04  K is above that.Ó 

{\it Carl Gibson: }
	ÒJust a comment.  As I understand, this is about 20 times the mass of anybody elseÕs estimate of the massive and electron neutrinos.  Is that right?Ó 

{\it Theo Nieuwenhuizen: }
	ÒYes.  It's completely ruled out by the WMAP analysis. They have 1\% of the neutrinos but they do assume cold dark matter there.  But this is replacing the cold dark matter.  It would be warm dark matter.  But if you assume we have cold dark matter you don't have much space for the neutrinos anymore.Ó

{\it Carl Gibson: }
	ÒAnd there's going to be an experimental test you had mentioned?Ó

{\it   Theo Nieuwenhuizen: }
	ÒOh yes.  It will be tested in 2015 I hope.  It will be the KATRIN experiment and they tell me it will start in 2015.  It's tritium decay in Karlsruhe, perhaps they start in 2012Ó.

{\it Francesco Sylos Labini:}
	ÒThank you.  IÕm not a fan of CDM at all.  Something that I've learned just by doing simulations and studying cosmology is that you need the matter to be cold in order to form structures.  If you're going to match hot dark matter how can you have some reasonable predictions on structure formation?Ó

{\it   Theo Nieuwenhuizen: }
	ÒThis we get the next item but it was Carl Gibson's talk in the morning.  Baryons don't need neutrinos or anything, they will do it by themselves. It's not even clustering but fragmentation what goes on.Ó  

{\it Rudy Schild: }
	ÒOther discussion of the neutrino and the particle?...OK, thank you.Ó 

\section*{SUBJECT 4: GRAVITATIONAL HYDRODYNAMICS}

{\it Theo Nieuwenhuizen: (Short intro)}
	ÒGreat.  We move on the next item which is Hydrogravitational Dynamics.  I have problems pronouncing that so I pronounce it as Gravitational Hydrodynamics, GHD.  And so, this stuff, what Carl gave his talk about, it goes in 3 steps.  The whole point is viscosity, 
	$\nu$, the viscosity, first in the plasma.  There you get the void formation.  You do viscosity and if you look a bit,  the red shift at which the viscous length scale enters the horizon happens to be $z=5,100$, well before the 1,100 of the decoupling.  So at the moment of decoupling the density is not homogeneous.  But, we have observed that the temperature is extremely homogeneous $10^{ -5}$ variations.  So there's something to explain there.  This void formation is a prediction.  This is step 1.  Step 2; we know all it's the Jeans mechanism which creates Jeans gas clumps of say 600,000 solar masses.  This we know since Jeans 1902. The idea is that all we have passed the decoupling ... so we have gas now and all the gas in the whole universe does what Mr. Jeans told us, it will break up in these clumps.   And then in each of these clumps ... there is again viscosity but now, not the viscosity of the plasma, this is $10^{26}$  m$^2$/s, but of the gas and this is $10^{13}$ m$^2$/s.   So this is much smaller, 13 orders of magnitude so you again get clumps but much, much less than what you had there and these clumps, they have earth mass and these have been discussed already a few times.  Which means that each of these Jeans clusters have something like $10^{11}$  of them. These earth mass H-He gas clouds are sometimes called micro brown dwarfs, today they were also called planets or small Jupiters.  Of these Jeans clusters in the Galaxy you have a million, $ 10^6$, in the Galaxy.  And in the Galaxy they are just the halo, so here you have your disk and they are everywhere.  They form an ideal gas in which they act as point particles.  These Jeans clusters act as point particles which traverse along the disk and along the milky way.  If they are point particles and if they are isothermal you have the flat rotation curves that come from the isothermal model.  You can find this in Binney and Tremaine. So you get the flattening of the rotation curves.  We have heard this item several times today.  And this is the end of my introduction.  So, are there any questions/remarks to this. 

{\it Alexander Unziger:}
	ÒHow do you form a halo?Ó

{\it Carl Gibson: }
	ÒThey can interact and things like to accrete to a central mass on a accretion disk.  I don't know how better to say it.  I mean if you just throw particles up into the air ... It's like Saturn.  It's more comfortable to come in on the plane of rotating massive object.  I'm sure there are people in this room that know how accretion disks are formed.  Stickiness helps. 

{\it Rudy Schild: }
	ÒCarl? Would the clump that was to become a galaxy have a net rotation which would enable the accretion disk I suppose?

{\it Carl Gibson: }
	ÒIt doesn't need it.  All it needs to do is have a mechanism to collide a little bit and you can actually see the orbits of  these things.  Somebody mentioned star streams around dim globular clusters.  The PAL, the Palomar Observatory.  Very dim globular ... Ó
	       
{\it Rudy Schild:  }
	ÒPAL 15.Ó

{\it Carl Gibson: }
	ÒYes.  This is a ... I mean all these PAL ... you know, very dim globular. This is why we're talking about this proto globular star clusters are just loaded with these planets.  Stars form at the edges because they get agitated from the planets and then you can actually see more stars in the wakes of these orbiting ... As Theo says, they orbit around the center of the galaxy and leave streams of stars to show they are there.  But they have this tendency ... they are trying to get into the center.  Some of them will diffuse far out and then they will go out to complex orbits and wind up in the plane of the galaxy.  The question comes:  where does the disk of a spiral galaxy come from?  Well, it's from these diffusing proto globular star clusters finding their way to this accretion disk.Ó 

{\it Theo Nieuwenhuizen: }
	ÒYou are now in a different paradigm.  As a paradigm you have two colliding galaxies, they merge and then half of the material ends up in the spiral disk. This is now gone because we have big halos.  And we have to start thinking from the halos.  Kroupa put up this picture.  We saw things that are explained in terms of halo behavior.Ó 

{\it Jan Smit:  }
	ÒI don't understand this viscosity. You would expect because the mean free path in the gas phase to be bigger that viscosity is bigger but, it's much smaller.Ó 

{\it Carl Gibson:  }
	ÒItÕs much smaller because the mean free bath is much smaller.  You have to understand the mechanism of viscosity in the plasma.  It's photon viscosity.  It's a photon colliding with an electron.  So, you have Thompson scattering, you know the cross section precisely and you know the density of electrons.  So you know precisely the mean free path but we also know the speed of the particle, the product is the viscosity.Ó 

{\it  Jan Smit:}
	ÒBut I thought the mean free path came with a denominator.Ó

{\it Carl Gibson: }
	Well, look at the dimensions.  It's the length times the speed, it is meter squared per second. That's the kinematic viscosity.  You're transporting momentum.Ó

{\it Jan Smit:  }
	ÒWhat is the role played in making these predictions?Ó

{\it Theo Nieuwenhuizen: }
	ÒAh ... that's the viscous length. Carl gave the formula for that.   If it is within your physical reach,
	within the sound horizon, you can get structures on that scale.

{\it David Wiltshire:}
	ÒWhenever you put something up, it's always important to think about what will falsify it and the questions that I would have would relate specifically to anisotropy spectrum in the CMB data because weÕre used to thinking of various features for example in terms of the ratio of baryonic matter to non-baryonic cold dark matter.  We usually think about the cold dark matter is clustered with some density contrast maybe $10^{-4}$  or something like this.  ItÕs more clumped than the baryons.  You'd be saying that the baryons are more clumped at some scale which has perhaps something to do with the planet size scale and zoning at the very high multiples.  Another issue ... IÕm not quite sure what part vorticity plays in this because vorticity perturbations are damped as long as we are expanding so the vorticity wouldn't have any effect on the large angles it would only at the very tiny angles.  Have you computed the CMB anisotropy spectrum or do you have some qualititative way of saying what features should be there?  Because, this calculation would surely be able to give you something that's predictable and could falsify it.Ó

{\it Theo Nieuwenhuizen: }
	ÒWhat we have is the position of the first peak and this is related to the size of the clumps you make here.  You will have voids and clumps.Ó 

{\it David Wiltshire:}
	ÒSurely the position of the first peak is just geometry and surface of last scattering.  It's got nothing to do with the small details..Ó 

{\it Carl Gibson:  }
	ÒIt has to do with the speed of sound in the plasma.Ó  

{\it David Wiltshire: }
	ÒCorrect.Ó

{\it Carl Gibson:  }
	ÒThe speed of sound in the plasma determines the speed of the rarefaction wave at the boundary of the proto super cluster voids.  See, you have a rarefaction wave ... Ó 

{\it David Wiltshire:}
	ÒWell, my understanding is that the first peak is a compression peak and there are two things involved; one is the speed of sound of the plasma and the distance of the surface of last scattering cause there's a degeneracy as people have been talking about.  So, just having the speed of sound by itself is not  enough.Ó

{\it Carl Gibson: }
	ÒThis produces a length scale at the surface of last scattering (as you put it) because, you know, your proto super cluster void fragments at $10^{12}$ seconds and then you have this sudden empty space formed that then everything surrounding this empty space falls away from the empty space.  So the void expands.  How fast does it expand?  You know this from gas dynamics or plasma dynamics it can go no faster than the speed of sound.  Mach 1 from thermodynamics.Ó 

{\it Theo Nieuwenhuizen: }
	ÒWhat Carl is saying is there is a candidate physical mechanism, but the precise calculation has to be done.  We are not at that level and we would kindly invite everybody interested. 

{\it David Wiltshire:}
	ÒSee I think , well at least I know that people, in Sydney actually, some people that are looking at back reaction who do these perturbative  things have put in viscosity and things like that and these thing are always very small.Ó

{\it  Carl Gibson: }
	ÒWhat number did they put in?Ó 

{\it David Wiltshire:}
	ÒAhhh ...  I don't know.Ó

{\it Theo Nieuwenhuizen: }
	ÒThere is the problem with viscosity.  Namely ... you can have nonlinear effects and that's a problem. 
	 If you don't put that in you don't get it out.Ó 

{\it David Wiltshire:}
	ÒBut we're talking about any part where the universe was very close to homogeneity. Ó

{\it Theo Nieuwenhuizen: }
	ÒBut there was fossil of turbulence and linear regions and they will trigger your stuff and those are never in models.Ó 

{\it David Wiltshire:}
	ÒOn very tiny scales ... I can imagine something but, I can't imagine big effects. Ó 

{\it Carl Gibson: }
	ÒYou have to understand that in this model you have a turbulent big bang, you know, full of Kolmogorov turbulent big bang at  scales  between $10^1$ and up to $10^8$.  So you have a spinning turbulent cylinder.  Think of it that way, with a gas of blunt particles and antiparticles.  Stretch  out the vortex lines ... it does turbulent mixing of the temperature and it is fossilized as soon as you get inflation.  You're beyond the scale of causal connections so it can't move.Ó 

{\it Theo Nieuwenhuizen: }
	ÒThank you.  Francesco.Ó

{\it Francesco Sylos Labini:}
	ÒSo maybe I've surely missed a lot of steps in these ideas but, so basically the thought is that you are including viscosity terms in a gravitational system.  But I don't understand exactly; what's the meaning with your isothermal model to explain reflecting of the rotation curves?Ó

{\it Theo Nieuwenhuizen: }
	ÒWell, given the fact that you have these planets,  what will they do?  Well they will ... you have somehow a Jeans cluster of these objects and what happens when these things inside the Jeans cluster?  Well, that will be the violent relaxation that you have already mentioned and therefore they go to the isothermal model.  So, inside the Jeans cluster you will have some rotation curves which are flattened but then you will have a million of these Jeans clusters in the galaxy and also these things among each other will thermalize  according to violent relaxation and therefore explain flattening of rotation curves of the galaxy.Ó 

{\it Francesco Sylos Labini:}
	ÒNo!  With violent relaxation never would you form an isotropic sphere, never.  I mean, why would you say that.  And then, something I do not understand.  You are talking about arguing no linear process of structural formation.  The description is not so ... Ó
 
{\it Theo Nieuwenhuizen: }
	ÒThere, we make two statements.  To let these things be there, you say they don't make the isothermal model.Ó 

{\it Francesco Sylos Labini:}
	ÒViolent relaxation doesn't give you an isotropic sphere.Ó
 
{\it Theo Nieuwenhuizen: }
	ÒThen what does give it to you.Ó 

{\it Francesco Sylos Labini:}
	ÒIt's another profile.  It depends on many things.  Whether the system is isolated or whether the system is not. This is well known also the arrows that people get in the simulations and they are not  ...    ...  clearly because they are not isolated and never told that  ...    ...   ...   ... It's many things.  It's a complex story. 

{\it Carl Gibson: }
	ÒYou forgot to tell Francesco, that these initially gaseous hydrogen clumps,  primordial gas planets, were gas planets.  Real gas planets, hot gas planets.  They gradually freeze and then they become collisionless and then what Theo says is right.  They first start off with the same density as the $10^{17}$ kg/m$^3$, very dense,  $10^4$ times more dense than the Galaxy.  But then they cool down with universe and every one of these fluffy gas clouds gets colder and colder and finally it freezes.  It gets to the freezing point of hydrogen which is 13.8 K.

{\it Rudy Schild: }
	ÒThat 13.8 degrees crosses the photon temperature when the universe is at redshift $z= 5-6$.Ó

{\it Theo Nieuwenhuizen: }
	ÒOK, but still, Francesco, your point is taken.  You say it does not go into the isothermal sphere.Ó

{\it Francesco Sylos Labini:}
	ÒViolent relaxation is for a collisionless system.  They provide a density that falls of as $1/r^4$.  
	For an isothermal sphere you need $1/r^2$.

{\it Theo Nieuwenhuizen: }
	ÒWeÕll come back to that.  Other points, other statements?  Ah, Richard, are you up.

\section*{SUBJECT 5: MISSING BARYONS, LENSING IN AN INHOMOGENEOUS UNIVERSE}

{\it Richard Lieu:}
 	ÒIt's to do with the point made by Tom Shanks this morning about the baryons.  The Coma cluster, this is actually a topic that by now, it's not a new topic, but by now it has about 200 citations.  For instance ... this particular paper is 2003 already and if you look at the Coma cluster in the normal weak  field, which is what Tom is talking about, the normal spatial extent is 1 degree in radius and that already harbors an awful lot of baryons.  If you look at it in soft X rays it goes out to 3 degrees.  That's sometimes the most extreme soft of excess.  More commonly known as the soft excess and if you look at an image of it, you see that the cluster extends to some enormous distance, about several mega parsecs.  This halo is actually a lot bigger than the normal X-ray halo.  In fact I think that the low level SZ in fact Plank detected in the form of that finger you saw the other day connecting three clusters may well be, in fact looking at another system of a manifestation of the same thing and that is that at the outskirts there are still baryons falling in.  Large amounts of them are not yet virialized and therefore at a lower temperature of about 1 million degrees, consistent with what Ostriker, in that paper I showed you the other day, predicted in 1998 that missing baryons, 50\% of them in the near universe, still in the form of this million degree temperature gas.  That's very, very difficult to observe.  So we still have to keep a close eye on the baryons when trying to solve cosmological problems. 

 The other point we may need to be making is connecting with the question of the back reaction and so on. This paper appeared in 2005 in which concerning the averaging to recover the FRW universe that is.  In the context of gravitational lensing we heard from Anthony Lewis yesterday.  There are issues even on lensing, not to mention the back reaction problem ... to do in an inhomogeneous universe.  When a universe has clumps and voids, how do you recover an average lensing magnification that equals the FRW universe of the same mean density?  It turns out, among many different ways doing the average magnification, only a particular rather esotheric way would give you back that consistency and itÕs known as the average reciprocal magnification.  Which basically means you take the 1 over the average of 1 over something and you call that the average of something.  And you do that ... then the next thing is whether there's any means to performing such an average and if it in particular corresponds to the magnification that's relevant to WMAP data.  So that's why I do strongly believe now that the question is not just simply the data pipeline of WMAP, that's part of the interpretation force but also, the model itself.  How do you implement the model?  Because you know that there are self consistency, very big problems, concerning what's the model predicting homogeneities at low red shift. Then how do you propagate the inhomogeneities back in terms of the construction of a mean Friedman universe.  You can't do it.  Then you just have to admit, even if you can, is your averaging process corresponding to reality in terms of whether it resonates with the observational matter.  ThatÕs all I wanted to say.  Thank you.  

{\it Theo Nieuwenhuizen: }
	ÒThank you.  Are there some remarks on this point?  Everybody agrees?  Richard, you have been very convincing.

{\it Richard Lieu:  }
	ÒWell, everybody is very, very tired.Ó

{\it Theo Nieuwenhuizen: }
	ÒNo, no, no ... they would throw tomatoes if they could.  Well, thank you Richard.  It is obvious we have a problem here and we don't know how to solve it.  

\section*{THE FINAL}

{\it Theo Nieuwenhuizen: }
Are there other things we would like to discuss?

{\it Alexander Unzicker: } 	
	ÒI would just like to draw the attention to a historic perspective.  We all know about the epicycles and we realize that nobody really likes the up to date model.  But what about in 20 or 50 years, how will science evolve.  I think there could be a good chance all people doubting today and having their serious problems with the model, just dying out and the dark matter and dark energy ...  all this stuff will be very accepted and the community will be ready to digest new parameters.  How can we avoid this ... 

{\it Theo Nieuwenhuizen: }
	ÒWe have an item here and that is the sociology of the community.  Of course, you being here and still at the last moment are not directly the people ... the people I need  to explain it.  We have already heard several people have had troubles getting their stuff published.  Although probably the main scientific approach was more or less correct.  It need not be completely correct.  What we can at least do is when we meet other people, try to open up their minds for a more fluent attitude concerning new ideas.  Because there are some problems we have seen them, but right or wrong, that is another issue.  But there are problems.  There are at least 20 problems with $\Lambda$CDM, so there maybe something out in any case and hopefully we get, maybe in 10 years weÕll have it done, we'll have a better model.  Still including or not including an $\Lambda$CDM, but a better theory and $\Lambda$CDM won't do this alone.  So new ingredients are needed but of course you are the wrong parish to preach for it, those that come in to church are supposed to believe as opposed to not believe, but if we can spread out the news to others they'll be a bit more open minded to accept papers and fund proposals in this direction and in such a way that the student is not immediately killed off after his papers but he/she  can continue his approach because what he did was sound physics whether or not the ballpark was correct in the end.  I think that's the right approach.  Yes, Luciano.Ó 

{\it Luciano Pietronero: }
	ÒMaybe I can make a little comment in the respect of statistical physics which is what dragged me in this business.  So if we take a fresh look at the situation of clustering, essentially what we observe is a fractal structure basically.  Whether there is homogeneity or not is not clear but for sure there is fractal structure for at least 2 decades.  As Michael Joyce pointed out, this is very interesting conceptually.  It's not power load one with respect to an explanation but there's a deep meaning in sense of self organized criticality and so on.  So in my opinion the simple issue, one which I believe can be resolved, is like, are these simulations good enough to produce a fractal structure, yes or no?  Do they produce it when you do them well?  I mean, basically,  the approach of Michael Joyce was very clean.  This is ... just this one piece ... you know, whether clustering is fractal or not is important in view of self organized criticality.  The other thing that ... I mean, many things I had, I would comment, but I don't.  The other thing that really surprises me a little bit is the little attention on the fact that after all the very large structures out there and there and they are very large, now, as I understand, the structure formation even with tons of dark matter and whatever people put, these structures so large and so are impossible to build from the point of decoupling from $z=1000$ to now with gravity.  So, I wonder, I mean, such a big structure of 300 to 400 mega parsecs with large amplitude, which is now being clear that it has a large amplitude, where does it come from?  Can we think of other mechanisms to build structures than gravity?  For example, a combination, some other dynamics.  So this should be considered.  So I would say, there is a simple part which is, I mean, well understanding gravity from other simulations but also a very open ended story of the very large structures.

{\it Carl Gibson: }
	ÒThe answer is no.  You need gravity.  But, gravity will do it. By making super cluster voids at the speed of sound in the plasma.  This will make you a void that is $10^{25}$ meters 10\% of the size of our horizon.Ó

{\it Luciano Pietronero:    }
	ÒIn how long of a time?  

{\it Carl Gibson: }
	ÒFrom then till now, $10 ^{12}$ seconds.Ó

{\it Luciano Pietronero:    }
	ÒFrom the decoupling?Ó

{\it Carl Gibson: }
	ÒFrom the big bang,  30,000 years.  So, it starts early, at the speed of light.

{\it Theo Nieuwenhuizen: }
	ÒOK, thank you.  The point has been made.

{\it David Wiltshire: }
	ÒI would like to ask a question.  If people ...   ...  these very large structures in numerical simulations may be  ...    ...  if general relativistic effects are put into the simulations? So, if these very large structures that seem to very difficult at finding $N$-body simulations are actually a result of general relativistic effects that aren't being included?Ó

{\it Alexander Unziger:}
	ÒSo what do you mean?Ó

{\it Carl Gibson:}
	ÒWell, hundreds of mega parsecs.Ó 

{\it David Wiltshire:}
	ÒWell, it's impossible to answer that question without having done the simulations.  The only person I know who claims they have done anything like this in his spare time is Voychek Ludman, I mean, he's never published anything so I don't know what he's done.  He just says he gets things happening a lot faster so ... Ó 

{\it Theo Nieuwenhuizen: }
	ÒAlexander ... would you like to answer to this point?Ó

{\it Alexander Unzicker:}
	ÒYes ... there are hints.  I mean, usually you can have really optimistic corrections here in the strong field.  Usually if you talk about evolution, because the weak fields are slower accelerations.  There is no hint that this could change things.Ó

{\it Ruth Durrer: }
	ÒThere are two places, the strong field and the large scale, this is large scale? Yes ... I thought so.Ó 

{\it Carl Gibson: }
	ÒWell, even Peebles has said that the size of the biggest hole or void or whatever you want to call it is the way that $\Lambda$ cold dark matter hierarchical clustering can be falsified.  If the voids get bigger than about  $10^{23}$ to $10^{24}$ meters he will give it up and now they're $10^{25}$.  So ten times  bigger.Ó

{\it Theo Nieuwenhuizen: }
	ÒHe will not give up.Ó

{\it Carl Gibson: }
	ÒWell, all we can do is argue with the data.Ó 

{\it Theo Nieuwenhuizen: }
	ÒOK, we got your point.  Richard.Ó

{\it Richard Lieu:}
	ÒFirst is does not depend on $\Lambda$ and even on cluster scales, a cluster mass of $10^{15}$ solar mass.  Then beyond the distance about 20 mega parsecs actually $\Lambda$ shields the gravitational force.  And in fact 2 clusters in other words, or a cluster and a galaxy satellite, their separated by 20 mega parsecs they will feel each otherÕs mutual attraction.  That's actually a neutral zone.  Beyond which it become repulsion in fact, and that's why there is this anti-correlation turning over that requires $\Lambda$.  That's something that you do need GR to produce.  That's almost like a barrier penetration. Ó 

{\it Theo Nieuwenhuizen: }
	ÒThis reminds me of a remark I have made before.  If these neutrinos are true then they will be free streaming for a long time and homogenous so the voids you then have are only baryonic voids which are still filled with a lot of neutrinos.  There is much more dark matter than baryons.  So basically you'll have a uniform universe still then but  later at redshift of 5  or 10 maybe, these neutrinos will condense on the baryons and then the voids will be really empty.  And then you should have strong back reaction effects because voids are empty and matter is clumped.Ó 

{\it Carl Gibson:  }
	ÒThey're totally empty of everything.  They should be.Ó 

{\it Rudy Schild: }
	ÒSo you would say that is a prediction, actually?Ó

{\it Carl Gibson: }
	ÒWell, exactly.  This is the difference between the $\Lambda$ cold dark matter hierarchical clustering model and this ...  I like to call it HGD, is that the last thing that  happens to the baryons,  you know, when you make super clusters by hierarchical clustering, in between those is the voids, but it's not empty.  It's still full of the neutrinos. However, when you start from fragmentation.Ó 

{\it Theo Nieuwenhuizen: }
	ÒThank you Carl.  Anna, we have not heard you.  Is there anything you would like to say or ask?

{\it Anna Ach\'ucarro:}
	One thing that I would have like to have asked this morning was, ... what observations would disprove your theory?

{\it Carl Gibson: }
	ÒWell if you ... I mean, what I was saying was that for every star in the galaxy there are 30 million earth mass Jupiter-like planets, not 8 or 9.  So, this is a rather strong prediction that is easily verifiable.  If one looks where I say they are and you don't find them then ... I mean, I think I have shown you pictures of them in the planetary nebula.  If those turn out not to be planets, if somebody manages to get there, and show that this is a Rayleigh Taylor instability in the gas then I will give it up.Ó 

{\it Theo Nieuwenhuizen: }
	ÒThank you Carl.  Rudy, you want to make a statement?Ó 

{\it Rudy Schild: }
	ÒSo, to answer your question, an important prediction of this idea as Carl is saying is the existence of all of these planets.  In Europe you've had two experiments looking for these but, I don't think they did a very good job.  They were basically biased to look for stellar mass objects and that comes in the cadence of making one observation per night.  And so, in that way, you would find stellar mass objects for which a micro lensing macho profile would take 10 or 20 days and this is seen in the Macho program in America, Eros and Ogle here in Europe.  There have been some attempts to repeat the experiment with a more rapid cadence but only with a small telescope and so they look at bright stars in the Magellanic clouds and therefore they were looking at super giants and there was no hope of seeing the micro lensing objects on large sources.  You have to have a point source and a point lens and so there were 2 assumptions in the attempts to preclude them.  Micro lensing of quasars gives a very different result.  Micro lensing of quasars where we look at a distant quasar through a foreground galaxy and the graininess in the foreground galaxy produces micro lensing with its character 6 signature the event duration telling the size.  This occurs at optical depth 1 in each gravitational lens, meaning, there would be continuous micro lensing by the baryonic dark matter or the galaxy dark matter.  It has shown, first by me actually in Q0957 and then 4 other observatories found the same signature in 5 or 6 different gravitational lenses, found that yes, they always show a highly probable optical depth $\tau = 1$ approximately signal indicating that, yes there is a $\tau = 1$ distribution of rogue planet mass not necessarily $10^6$, sometimes $10^5$, sometimes $10^4$  objects.  We have now, a new experiment to repeat this. We now have a sense of where to look in the Magellanic clouds because we believe that we have now seen the thermal signature of the baryonic dark matter.  Now Tom, listen ... I didn't get a chance to describe this to you.Ó

{\it Theo Nieuwenhuizen: }
	ÒCarry on ... Tom is on board.Ó 

{\it Rudy Schild: }
	ÒIf Carl is right, the planets would have been made everywhere in the halo of the galaxy in these clumps.  These clumps have now been seen in these very low temperature maps like Dirbe at 200 microns and the longest WMAP and in some other experiments and these have given a spectral signature by showing a peak at a temperature always prone to be about 15 degrees  Kelvin.  That's seen in the halo of our galaxy in these globs that they see and also halos of other galaxies.  So, why is there a spectral signature at 13.7 K?  It is because the triple point of hydrogen at near vacuum conditions is at 13.8 K.  So these objects are cooling but they get stuck at the triple point because of the heat of fusion and so we see them everywhere and we call them infrared cirrus.  By looking at the infrared cirrus clumps seen in the maps of the Magellanic clouds and doing the micro lensing experiment, the same Macho, Ogle, Eros, experiment through the Magellanic clouds through these 13.8 degree clumps which should then be the hydrogen spheres, we would then be to detect whether, in fact, that prediction of the theory was correct.  Those should be the hydrogen spheres.Ó                                                                       

{\it Theo Nieuwenhuizen: }
	ÒAnd if they are not there ... then the game is over.Ó 

{\it Rudy Schild: }
	ÒIf they're not there ... then I shut up.Ó

	{\it Anna Ach\'ucarro:}
	ÒI guess what I was asking, from what I understood, the micro lensing ... you can detect these things obviously.  In your history, it would seem to me that your combination was more or less ... I should expect to see a micro cosmic background ...  your peaks are coming from somewhere else.

{\it Theo Nieuwenhuizen: }
	ÒOK, let me take it from Carl.  This is a weak point because we didn't work it out and we are not in that field.  We would invite people to do it for us.  The idea is that the plasma breaks up into clumps and the size of the clumps is more or less the position of the first peak.
	 Now the higher peaks, they will come from inside these clumps and this is, I think, more or less what it is now.  It will be gas dynamics more or less as it is done now but there will not be dark matter for us so, thereÕs a good chance that it will fail because we don't have dark matter, as these neutrinos do nothing.  They are just homogenous, they do not help and baryons must do it alone.Ó

	{\it Anna Ach\'ucarro:}
	ÒOK, so you will be in a position to make a prediction that you can compare?Ó 

{\it Theo Nieuwenhuizen: }
	ÒIf this works out, it should be compared to the CMB, now, listen ... if you have a real theory you must describe everything.  Including all observations that are well established.Ó

	{\it Anna Ach\'ucarro:}
	ÒI also had a question about the Bullet cluster but I don't know whether you have discussed it or not.Ó 

{\it Tom Shanks: }
	ÒThe Bullet cluster is often used ...    ...  for a cold dark matter model.  At least the first paper, I thought, was a swindle.  You know what I'm talking about?  I looks like Bullet 2 clusters and that the gas in the galaxies is separated out and then you ask the question: where the gas is.  Is it in the galaxies?  If it's something gas, then it could be a baryonic model or MOND model.  If it's in the galaxies, that's where you would expect the cold dark matter to be, as collisionless particles, associated with the galaxies around the gas.  And so, the gravitational lensing results seem to favor the idea that the mass was with the galaxies, and therefore the cold dark matter rather than the gas.  But, we actually took very nice pictures that looks like the gas is very separate, separated out in this cluster by the event of the one cluster passing through another.  We actually get numbers, the gas is where they say the gas is quite spread,    ... as is the mass from lensing so the only way they got the result was by subtracting off a slender circally  symmetric profile set it on the galaxy position which then means that you see no mass near the gas.  But if they had done it the other way around they would have gotten the completely opposite result.  So that's why I say it's somewhat of a swindle.  It depends on you thinking that this structure, the gas is separated out from the galaxies when actually it hasn't been separated out long enough to do the experiment.  So I don't think that is a very good argument for cold dark matter down to the masses for cold dark matter cluster.  When you begin to look at the details ... Ó     

{\it Theo Nieuwenhuizen:}
	ÒWell now, I think we have almost set up everything that we knew before we came here.  We have put it upside down.  Some things have been said back again.  Other things are still  ...   in the air.  Alexander, you had a nice quote, can you read it?Ó

{\it Alexander Unzicker: }
	ÒI will not talk, I will just quote and shut up and not comment.  It's Richard Feynman, and he said ``Correct theories of physics are perfect things, and the replacement theory must be a new perfect thing, not an imperfection added on to an old perfect thing.''  This is the essence of revolution the replacement of the old with the new.  Not the adding of more crap on to the old.Ó

{\it Theo Nieuwenhuizen:}
	ÒWell, for me this is the end.  
We have already thanked the Lorentz Center, I would still like to thank my co-organizers.  First of all Ruth; Ruth, you came in late, we were very happy you were willing to do that, thank you. (applause) Rudy, you are our ``Nestor''.  You know the best.  You have all the experience, thank you very much. (applause)  But I think you all know who was the real organizer of this meeting, Francesco, thank you. (applause)  

{\it Rudy Schild: }
	ÒAnd someone has to thank you Theo.Ó (applause)

{\it Theo Nieuwenhuizen:} 	ÒOK,  it's  done.Ó

\section*{Acknwledgement}
The authors are most grateful to Mrs. Sue Verbiest for accurately typesetting the text from the audio recording.

\end{document}